\documentclass[doublecol]{epl2} 
\usepackage{hyperref}
\usepackage{graphicx}

\usepackage{color}
\usepackage[normalem]{ulem} % gives command \sout

\usepackage{xcolor}
\definecolor{shadecolor}{RGB}{255,255,255}

\usepackage{comment}
\newcommand\rmd{{\rm d}}

\newcommand\Qdot{J}

\renewcommand\phi{\varphi}
\renewcommand\rho{\varrho}

\definecolor{DarkGreen}{rgb}{0,0.4,0}

\renewcommand{\vec}[1]{\mathbf{#1}}

\begin{document}

\title{Comment on
``Inconsistency of the conventional theory of superconductivity''
by J. E. Hirsch}

\author{
Denis M. Basko%\footnote{\ denis.basko@lpmmc.cnrs.fr}
\inst{1} \and
Robert S. Whitney%\footnote{\ robert.whitney@lpmmc.cnrs.fr}
\inst{1}
}
\shortauthor{D. M. Basko and R. S. Whitney}

\institute{     
\inst{1} %Laboratoire de Physique et Mod\'elisation des Milieux Condens\'es, 
Universit\'e Grenoble Alpes and CNRS, LPMMC,
% \phantom{\inst{1}}
25 Avenue des Martyrs, BP 166, 38042 Grenoble, France
 }
\date{June 22, 2020}


% == Following command is hack to remove words "EPL DRAFT" from header of first page 
%           It works by screwing-up the style of 1st page.  The big problem is that it starts the text 
%           with "eplplaineplfirst", so I have hidden  this text behind a colored box containing the date
\renewcommand\thispagestyle{eplplain}

\abstract{
J. E. Hirsch [EPL {\bf 130} (2020) 17006] 
claimed an inconsistency between thermodynamics and the theory of superconductivity. 
We argue that he overlooked a crucial term which determines the supercurrent dynamics and ensures energy conservation by providing an internal energy source for the Joule heating.
Thermodynamic consistency is restored by restoring energy conservation. 
The correct dynamics is given by Maxwell's equations in the superconductor.
}

\maketitle

%===========================

%%%%%%%%%%%%
% Following command makes a white frame (since black!0 means black at 0%) moved to hide text "eplplaineplfirst" which is due to my hack of style. 
\vskip -5truemm  
\noindent\colorbox{shadecolor}
{\parbox{\dimexpr\columnwidth-2\fboxsep\relax}{$\phantom{|}$}}
\vskip -12truemm
$\ $
%%%%%%%%%%%%

\section{Introduction}
J. E. Hirsch \cite{Hirsch,Hirsch2} considered a thought experiment with a superconductor in an external magnetic field cooled via a thermal contact to a cold reservoir. 
The superconductor is treated as two fluids 
--  a normal quasiparticle fluid  and a condensate (Cooper-pair fluid) with a time-dependent London penetration depth~$\lambda_L$.
Ref.~\cite{Hirsch} identified internal (Joule) heating of the normal fluid, when fast cooling 
induces a rapid change of~$\lambda_L$. 
Ref.~\cite{Hirsch} then found that the entropy change depends on the cooling rate, which is inconsistent with the fact that entropy is a state function (so its change is given by the initial and final states and nothing else).

By assuming  that $\lambda_L$ entirely parametrized the problem, with Joule heating induced by its rate of change,
Refs.~\cite{Hirsch,Hirsch2} implicitly assumed $\lambda_L$ was a thermodynamic displacement. 
However, this displacement was inconsistently treated, Refs.~\cite{Hirsch,Hirsch2} overlooked the conjugate force that gives $\lambda_L$  energy-conserving dynamics in which it deviates from its instantaneous equilibrium value. Thermodynamic consistency is restored by these dynamics.

A second issue is Ref.~\cite{Hirsch}'s assumption that the magnetic field retains an equilibrium shape when cooled fast, so $\lambda_L$ is the only parameter.  Using the superconductor's free energy and Maxwell's equations, we find this to be wrong; the vector potential ${\bf A}({\bf r})$ changes shape away from equilibrium.  Thus the relevant thermodynamic displacement is the whole field ${\bf A}({\bf r})$ rather than a single parameter, $\lambda_L$.

%%%%%%%%%%%%%%%%%%%%%%%%%%%%
\begin{figure}
\centerline{\includegraphics[width=0.8\columnwidth]{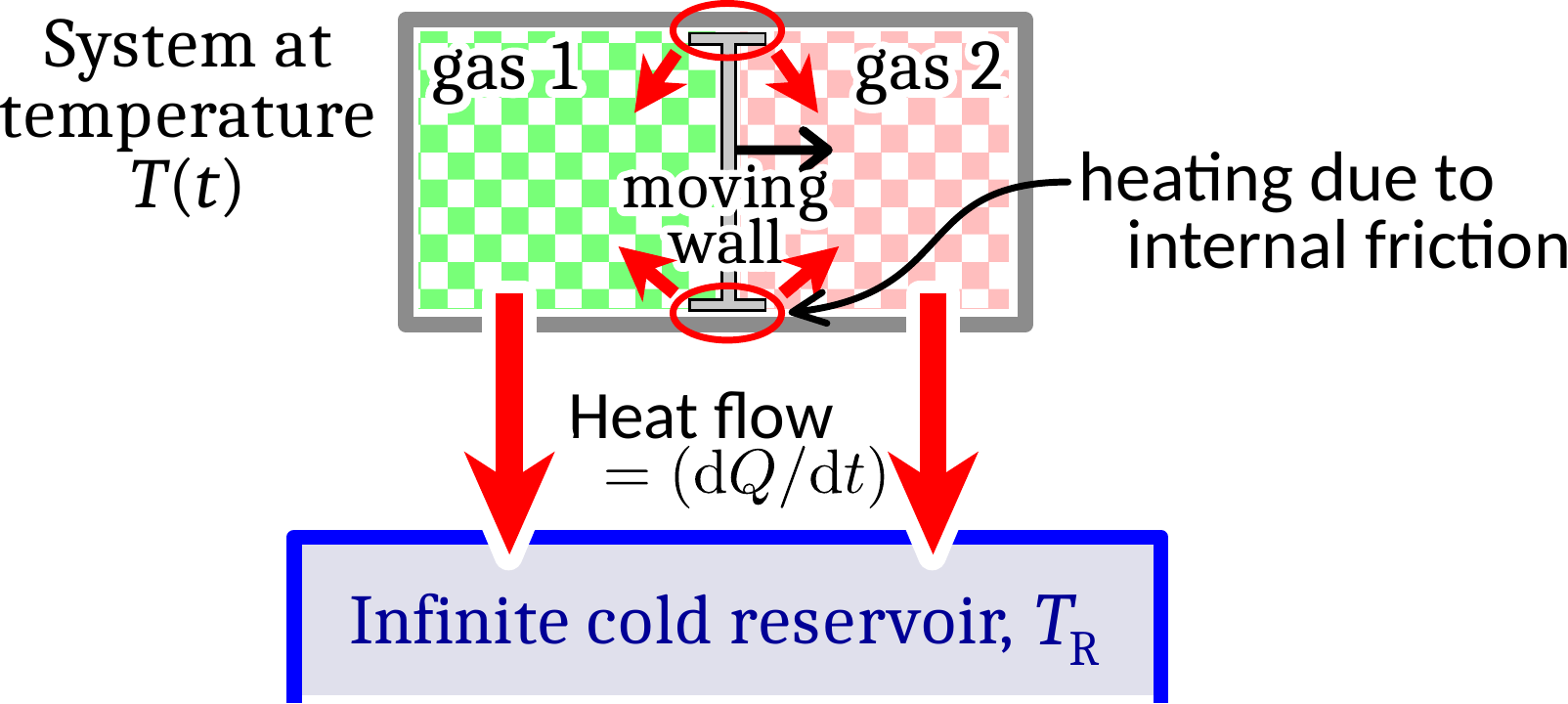}}
\caption{
\label{Fig:2-gas-piston}
A simpler system exhibiting internal heating as it cools.
A closed cylinder containing two gases separated by a moving wall (position $x$) which transfers heat perfectly. Gas 1 is ideal, and gas 2 has attractive interactions (Van der Waals gas); their pressures are $p_1$ and $p_2$. Upon cooling, $p_1$ drops more slowly than $p_2$ (due to the interactions), and the wall moves to the right, because $x$ has a conjugate force  $f=(p_1-p_2)$. Then Eq.~(\ref{eq:relaxation}) is Newton's second law for the wall with zero mass and a viscous friction force $-\eta\ (\rmd{x}/\rmd{t})$.
}
\end{figure}
%%%%%%%%%%%%%%%%%%%%%%%%%%

\section{Energy conservation}
A proper description of internal relaxation and heating involves an internal thermodynamic displacement~$x$, being pushed towards its equilibrium value $x_\mathrm{eq}(T)$
by its conjugate thermodynamic force $f(x,T)=-(\partial{F}/\partial{x})_T$.
Here $F(x,T)$ is the system's Helmholtz free energy (we assume constant volume and omit it); see Fig.~\ref{Fig:2-gas-piston} for an example.
At equilibrium $f\big(x_{\rm eq}(T),T\big)=0$, since $F(x,T)$ has a minimum with respect to~$x$.
%Thus $f$ is crucial in determining the system's trajectory $\big(x(t),T(t)\big)$ as a function of time $t$.
The work done by the system when $x$ changes is $f\,\rmd{x}$. 
The system's entropy $S(x,T)=-(\partial{F}/\partial{T})_x$.

Now let there be internal (Joule) heating
$\rmd{Q}_J=\eta(x,T)\, (\rmd{x}/\rmd{t})\,\rmd{x}$, 
with  $\eta(x,T)$ being  a kinetic coefficient given by the system's microscopics.
The energy source for this is the work $f \,\rmd{x}$, so energy conservation gives
\begin{equation}
\label{eq:relaxation}
\eta(x,T)\,(\rmd{x}/\rmd{t}) = f(x,T),
\end{equation}
which describes relaxation of $x$ to $x_{eq}$ \cite{Landafshitz6}.
The first law gives the system's internal energy change
\begin{equation}
\rmd{U}=(-\rmd{Q}+\rmd{Q}_J)-f \,\rmd{x}=-\rmd{Q}, 
\label{Eq:dU}
\end{equation}
where $\rmd{Q}$ is the heat flow into the reservoir.
Ref.~\cite{Hirsch} overlooked $f \, \rmd{x}$, 
leading to Joule heating without an energy source.

\section{Entropy change with fast cooling}
At initial time $t_i$ the system is 
in equilibrium at $T(t_i)=T_i$. 
For Ref.~\cite{Hirsch}'s fast cooling protocol $(b)$, 
it is connected at $t_i$ to an infinite reservoir with temperature $T_2<T_i$, extracting heat at a rate $\Qdot$. It is disconnected at $t=t_i+\Delta{t}$, after heat 
$\Delta{Q}=\int_{t_i}^{t_i+\Delta{t}} \Qdot\,\rmd t$ has flowed into the reservoir.
Following Ref.~\cite{Hirsch}'s logic up to its Eq.~(24), we find the entropy change for arbitrary $\Delta{t}$ and any  final time $t_f \geq t_i+\Delta{t}$, 
\begin{eqnarray}
\label{Eq:DS^b}
\Delta S_{univ}^{(b)} = {\Delta Q \over T_2}-\int_{t_i}^{t_i+\Delta{t}} {\rmd t \over T}\, \Qdot +  \int_{t_i}^{t_f} {\rmd t \over T}\, \eta\left(\frac{\rmd{x}}{\rmd{t}}\right)^2.
\end{eqnarray}
The first term is the reservoir's entropy change, 
the second term is the system's entropy change due to heat flow to the reservoir, 
and the third term, $\Delta{S}_J$,  is due to internal (Joule) heating.
Let us use Eq.~(\ref{Eq:dU}) to write $\rmd{U}/\rmd{t}=-\Qdot$, and use Eq.~(\ref{eq:relaxation}) to eliminate $\eta(x,T)$.
Then 
\begin{eqnarray}
\label{Eq:DS^b-2ndresult}
\Delta S_{univ}^{(b)}= {\Delta Q \over T_2}+\int_{t_i}^{t_f} {\rmd t \over T}\, \left(\frac{\rmd{U}}{\rmd{t}} +  f\,\frac{\rmd{x}}{\rmd{t}}\right).
\end{eqnarray}
As  $\rmd{U}=-f\,\rmd{x}+T\,\rmd{S}$, the integrand is $\rmd{S}/\rmd{t}$, so the integral is $S(x_f,T_f)-S(x_i,T_i)$.
Thus $\Delta S_{univ}^{(b)}$ depends on the initial and final states, but not on the trajectory between these states, so we recover thermodynamic consistency.  If $x_i=x_\mathrm{eq}(T_i)$ and $x_f=x_\mathrm{eq}(T_f)$, then $\Delta S_{univ}^{(b)}$ is the same as for adiabatic evolution with the same change of $U$, so
$\Delta S_{univ}^{(b)}=\Delta S_{univ}^{(a)}$  in Ref.~\cite{Hirsch}'s language. 

\section{Short time Joule heating}
Ref.~\cite{Hirsch} erroneously estimated that $\Delta{S}_J = \mathcal{O}(\Delta{T})$ for short $\Delta{t}$ (small temperature change $\Delta{T}$), because it overlooked the dynamics of $\lambda_L$ and assumed it always took its equilibrium value. In our language it assumed $x(t) = x_\mathrm{eq}\big(T(t)\big)$ at all $t$. This is wrong because Joule heating implies damping that slows the dynamics of $x$, so that $\big|\rmd{x}/\rmd{t}\big| < \big|(\rmd{x}_\mathrm{eq}/\rmd{T}) (\rmd{T}/\rmd{t})\big|$ in contradiction with  Ref.~\cite{Hirsch}'s Eq.~(20).

Instead one must construct the trajectory $x(t),T(t)$ for protocol~$(b)$ by integrating Eq.~(\ref{eq:relaxation}) together with $\rmd{U}/\rmd{t}=-\Qdot$, linearised around the initial equilibrium~\cite{Footnote:supplementary}.
When the system and reservoir are disconnected at time $t_i+\Delta{t}$, one has $x(t_i+\Delta{t})=x(t_i)+\mathcal{O}(\Delta{t}^2)$ only, because $f=0$ initially. Subsequently, $x$ relaxes to its new equilibrium value $x_\mathrm{eq}(T_i-\Delta{T})= x_\mathrm{eq}(T_i)+\mathcal{O}(\Delta{T})$. This relaxation is given by Eq.~(\ref{eq:relaxation}) and ${\rm d}U/{\rm d}t =0$;
it occurs on an intrinsic time scale, independent of $\Delta{T}$. Eq.~(\ref{eq:relaxation}) ensures that $\rmd{x}/\rmd{t}=\mathcal{O}(\Delta{T})$ at all times, so the contributions to $\Delta{S}_J$ from $t_i<t<t_i+\Delta{t}$ and $t>t_i+\Delta{t}$ are $\mathcal{O}(\Delta{T}^3)$ and $\mathcal{O}(\Delta{T}^2)$, respectively.
Thus $\Delta{S}_J$ should be absent from Ref.~\cite{Hirsch}'s  
${\cal O}(\Delta{T})$ calculation, removing its inconsistency.

\section{Cooling a superconductor}
Our above results for generic $x$ and $f$ apply equally to the model in Ref.~\cite{Hirsch} or our Fig.~\ref{Fig:2-gas-piston}.  
Now we turn specifically to a type I superconductor, where the order-parameter $\Delta$ has a constant phase.
Take the  Ginzburg-Landau free energy, $F$, far enough from the critical temperature 
so that $|\Delta|$ relaxes so quickly that it is always in equilibrium \cite{Nagaosa}
(this assumption is relaxed in \cite{Footnote:supplementary,Kopnin}).
Then
\begin{equation}\label{eq:GinzburgLandau}
F[\vec{A},T]=\int{d}^3\vec{r}\left[\frac{\vec{A}^2}{8\pi\lambda_{eq}^2(T)}
+\frac{[\nabla\times\vec{A}]^2}{8\pi}+\mathcal{F}_0(T)\right],
\end{equation}
where $\lambda_{eq}(T)$ is the {\it equilibrium} London penetration depth  at temperature $T$,
and $\mathcal{F}_0(T)$ is the free energy density in the absence of currents and fields. 
The supercurrent $\vec{j}_s=-\vec{A}c/[4\pi\lambda_{eq}^2(T)]$
(assuming the local London limit, as in Ref.~\cite{Hirsch}).
The normal quasiparticle current $\vec{j}_n=-(1/c)\,\sigma(T)\,\partial\vec{A}/\partial{t}$, where $\sigma$ is the dissipative quasiparticle conductivity. 
$\vec{A}({\bf r})$, which includes both the vector potential of the external field and that of the currents, plays the role of the thermodynamic displacement.

%We take $\vec{A}({\bf r})$ as the relevant thermodynamic displacement, which includes both the vector potential of the external field and that of the currents.

To describe the dynamics, we plug the currents $\vec{j}_s,\vec{j}_n$ into the relevant Maxwell equation (Ampere's law). We neglect the displacement current, responsible for the emission of electromagnetic waves, as in Ref.~\cite{Hirsch}. Then
\begin{equation}\label{eq:Maxwell3}
\frac{\sigma(T)}{c^2}\,\frac{\partial\vec{A}}{\partial{t}}=
-\frac{\vec{A}}{4\pi\lambda_{eq}^2(T)}-\frac{\nabla\times\nabla\times\vec{A}}{4\pi},
\label{Eq:Maxwell3}
\end{equation}
in full analogy to Eq.~(\ref{eq:relaxation}). % with $x$ replaced by the field $\vec{A}(\vec{r})$.  
The right-hand side equals $-(\delta{F}/\delta\vec{A})_T$, so it is the {\it conjugate force}.  
If one were to assume a fixed spatial profile of ${\bf A}({\bf r})$ depending only on $\lambda_L(t)$ 
[as in Ref.~\cite{Hirsch}'s Eq.~(1)], then our Eq.~(\ref{Eq:Maxwell3}) would give an equation for $(d\lambda_L/dt)$ of the form in our Eq.~(\ref{eq:relaxation}).  This would restore thermodynamic consistency through its energy-conserving  dynamics, for which  $\lambda_L(t) \neq \lambda_{eq}(T(t))$.
Sadly, this ``fixed profile'' assumption over-simplifies the superconductor's dynamics, as seen below.

For small $\Delta{T}$,  the Joule heating while the system is connected to the reservoir is $\mathcal{O}(\Delta{T}^3)$, since the right-hand side of Eq.~(\ref{eq:Maxwell3}) vanishes in the initial equilibrium state. For the subsequent  relaxation, one can expand $\vec{A}(\vec{r})$ in eigenfunctions of the radial Laplacian. Each eigenfunction has its own relaxation time $\sim4\pi\sigma{R}^2/c^2$, independent of~$\Delta{T}$, 
so the field profile changes with $t$, invalidating the ``fixed profile'' assumption. 
Again, $\Delta{S}_J=\mathcal{O}(\Delta{T}^2)$~\cite{Footnote:supplementary}, contrary to Ref.~\cite{Hirsch}'s estimate.

%%%%%%%%%%%%%%%%%%%%%%%%%%%%%%%%%%%%%%%
%%%%%%%%%%%%%%%%%%%%%%%%%%%%%%%%%%%%%%%

\setcounter{equation}{0}
\renewcommand{\theequation}{S.\arabic{equation}}

\clearpage
\centerline{{\bf SUPPLEMENTARY MATERIAL}}
\vskip 3mm

Part 1 gives details on the short-time calculation outlined in our
section ``Short time Joule heating'' and at the end of section ``Cooling a superconductor''. 
Part 2 shows how to maintain thermodynamic consistency  near the critical temperature, where the approximation in our section ``Cooling a superconductor'' is not applicable. 

\vskip 7mm
\centerline{{\bf PART 1: FAST COOLING TRAJECTORY}}
\vskip 3mm

Here we derive the system's trajectory for Ref.~\cite{Hirsch}'s fast cooling protocol $(b)$.
To keep the analysis general, we consider an arbitrary thermodynamic displacement $x$ and conjugate force $f$.

Suppose one were to assume that  the superconductor has a fixed spatial profile of the magnetic field, entirely determined by the parameter $\lambda_L(t)$, as in Ref.~\cite{Hirsch}'s Eq.~(1).
Then one could get a  thermodynamically consistent theory, by replacing $x$ with $\lambda_L$ in what follows.   However, we have argued that this assumption over-simplifies the dynamics, because the spatial profile is {\it not} fixed for fast cooling.
Thus to have thermodynamic consistency {\it and} the correct dynamics, 
one needs to replace $x$ by the vector potential field $\vec{A}(\vec{r})$ in what follows.

We follow Ref.~\cite{Hirsch} in assuming the cooling occurs during a short enough time
(small enough change of $T$) that the system does not deviate far from its initial equilibrium state.
We assume the system to start in equilibrium at temperature $T_i$ for which the equilibrium displacement is $x_i\equiv x_{\rm eq}(T_i)$, and that the system stays close enough to this initial equilibrium state that we can expand  all quantities about it. 
We note that to get the system's equations of motion to first order in deviations 
from the initial equilibrium state, the free energy must be expanded to second order.

Expanding the equilibrium free energy  gives
\begin{eqnarray}
F(x_{\rm eq}(T),T)&=& F_i \,-\,S_i\,(T-T_i)  - \frac{C_\mathrm{eq}}{2T_i}\,(T-T_i)^2 
\nonumber \\
& & \qquad + \mathcal{O}\big((T-T_i)^3\big) \, . 
\end{eqnarray}
It will be seen shortly in Eq.~(\ref{eq:UxT=}), that the second-order's coefficient  $C_\mathrm{eq}$ is the equilibrium heat capacity at $T_i$ defined as $C_\mathrm{eq}(T)= \rmd U(x_{\rm eq}(T),T)\big/\rmd{T}$.
At the same time we expand $x_{\rm eq}(T)$ about $x_i$ as
\begin{equation}
x_{\rm eq}(T) = x_i + \alpha \,\big(T-T_i\big) + \mathcal{O}\big((T-T_i)^2\big).
\label{Eq:x_eq}
\end{equation}

To deal with  small deviations from equilibrium, we  
recall that the force is zero at equilibrium, so we treat it to linear order as
\begin{equation}
f(x, T) = -k(T ) \,\big(x- x_{\rm eq} (T)\big) + \mathcal{O}\big((x-x_{\rm eq}(T))^2 \big).
\end{equation}
Thus the free energy is quadratic in $\big(x- x_{\rm eq} (T)\big)$,
so it reads
\begin{eqnarray}
F(x,T)&=&F_i-S_i(T-T_i)-\frac{C_\mathrm{eq}}{2T_i}\,(T-T_i)^2
\nonumber \\
& & + \frac{k_i}2\left[x-x_i-\alpha(T-T_i)\right]^2,
\end{eqnarray}
where the subscript $i$ indicates that a quantity is evaluated at $T_i$; for example $k_i\equiv k(T_i)$.
From this we deduce the entropy, $S=-(\partial{F}/\partial{T})_x$ and the internal energy $U=F+TS$.
To linear order they read
\begin{eqnarray}
S(x,T)&=&S_i+\frac{C_\mathrm{eq}}{T_i}\,(T-T_i)
\nonumber \\
& & {}+\alpha{k}_i\,[x-x_i-\alpha(T-T_i)],
\\
U(x,T)&=&U_i+C_\mathrm{eq}(T-T_i)
\nonumber\\
& &{}+k_i\alpha{T}_i\left[x-x_i-\alpha(T-T_i)\right].\label{eq:UxT=}
\end{eqnarray}
% \begin{eqnarray}
% S(x,T)&=&S_i+\frac{C_\mathrm{eq}}{T_i}\,(T-T_i)
% \nonumber \\
% & & +\alpha{k}_i\,[x-x_i-\alpha(T-T_i)],
% \\
% U(x,T)&=&U_i+C_\mathrm{eq}(T-T_i)+\frac{C_\mathrm{eq}}{2T_i}\,(T-T_i)^2
% \nonumber\\
% & &+k_i\alpha{T}_i\left[x-x_i-\alpha(T-T_i)\right]
% \nonumber \\
% & &+\frac{k_i}2 (x-x_i)^2 -\frac{\alpha^2k_i}{2}(T-T_i)^2.
% \end{eqnarray}

We will find that the dynamics of fast cooling of the system has two parts.
The first part is the evolution, between $t_i$ and $t_i+\Delta{t}$, during which the system gives heat $\Delta{Q}$ to the reservoir.  This first part of the evolution leaves the system in a non-equilibrium state at time $t_i+\Delta{t}$;
a state with finite $\rmd{x}/\rmd{t}$.  If one then disconnects the system from the reservoir at time $t_i+\Delta{t}$
(so no more heat can flow to the reservoir), then the second part of the dynamics  occurs in which the system relaxes to equilibrium. In this second part of the dynamics, $\rmd{x}/\rmd{t}$ decays slowly to zero, and continues to provide internal (Joule) heating of the system as it decays.

\section{Trajectory while coupled to reservoir}
The system's trajectory is determined by Eqs.~(\ref{eq:relaxation}), (\ref{Eq:dU}).
For the first part of the evolution (between $t_i$ and $t_i+\Delta{t}$) they give
\begin{eqnarray}
\eta_i\,\frac{\rmd{x}}{\rmd{t}}&=&-k_i\left[x-x_i-\alpha\,(T-T_i)\right],\label{eq:diff}
\label{Eq:motion1}
\\
-\Qdot\, (t-t_i) &=&C_\mathrm{eq}\,(T-T_i)
\nonumber \\
& & +k_i\alpha{T}_i\left[x-x_i-\alpha\,(T-T_i)\right] .\qquad \label{eq:energy}
\label{Eq:motion2}
\end{eqnarray}
Since we are interested in small temperature changes, we have dropped quadratic terms, hence we can treat the rate of heat flow into the reservoir, $\Qdot$, as a constant between  $t_i$ and $t_i+\Delta{t}$.  
Then to get the dynamics when an amount of heat $\Delta{Q}$ is transferred to the reservoir, we need to integrate up to $t_i+\Delta{t}$ where $\Delta{t}\equiv\Delta{Q}/\Qdot$.

The solution of Eqs.~(\ref{Eq:motion1}), (\ref{Eq:motion2}) for $x$ is
\begin{eqnarray}
x(t)-x_i&=&\frac{\alpha}{C_\mathrm{eq}}\,\Qdot\tau\left(1-\frac{t-t_i}\tau -e^{-(t-t_i)/\tau}\right),\qquad
\end{eqnarray}
where we define  
$\tau=(\eta_i/k_i)\left[ 1\,-\,k_i\alpha^2T_i/C_\mathrm{eq}\right]$.
However as we dropped the quadratic terms from Eqs.~(\ref{Eq:motion1}) and (\ref{Eq:motion2}), this solution only holds at
lowest order in $t-t_i$, so
\begin{eqnarray}
x(t) -x_i \approx-\frac{\alpha}{C_\mathrm{eq}}\frac{\Qdot\,(t-t_i)^2}{2\tau}
+\mathcal{O}\big((t-t_i)^3\big).
\label{Eq:x_early-times}
\end{eqnarray}
Since  $x-x_i=\mathcal{O}\big((t-t_i)^2\big)$, one can set $x=x_i$ in Eq.~(\ref{Eq:motion2}) and find the temperature:
\begin{equation}
T(t)-T_i=-\frac{\Qdot\, (t-t_i)}{C_\mathrm{eq}-k_i\alpha^2T_i}
+\mathcal{O}\big((t-t_i)^2\big).
\end{equation}
Note the appearance of the heat capacity at constant~$x$, $C_x=(\partial{U}/\partial{T})_x=C_\mathrm{eq}-k_i\alpha^2T_i<C_\mathrm{eq}$ which was correctly guessed in Ref.~\cite{Hirsch}, whose Eq.~(17) is, nevertheless, wrong, because $\Delta{Q}_J={\cal O}(\Delta{T}^2)$, as we will see shortly.
Note that when the system is disconnected from the reservoir at time $t_i+\Delta{t}$, it has  $(d x/\rmd{t}) \neq 0$, so it is clearly {\it not in equilibrium}.  In addition 
$T_i-T(t_i+\Delta{t})=\Delta{Q}/C_x>\Delta{T}=\Delta{Q}/C_\mathrm{eq}$, 
which means that the system is overcooled at time $t_i+\Delta{t}$.

The situation would be fully analogous if we had the field $\vec{A}(\vec{r})$ instead of~$x$: the temperature change on this first stage would be determined by $C_\vec{A}=(\partial{U}/\partial{T})_\vec{A}$, while $\vec{A}(\vec{r})$ itself would change to $\mathcal{O}(\Delta{T}^2)$ only.

\section{Trajectory after decoupling from reservoir}
Now we consider the second part of the dynamics,
when the isolated system equilibrates after having been decoupled from
the reservoir at time  $t_i+\Delta{t}$.  Physically this corresponds to $(d x/\rmd{t})$ decaying  to zero,
causing internal heating of the system as it decays. This continued heating will compensate for the overcooling during the time the system was coupled to the reservoir.   
Since the heat that had flowed to the reservoir by time $t_i+\Delta{t}$ is $\Delta Q$, and this does not change during this part of the evolution, 
the system dynamics are given by Eqs.~(\ref{eq:diff}), (\ref{eq:energy}) with $-\Qdot\, (t-t_i)$ replaced by $-\Delta{Q}$.  Solving these equations gives
\begin{eqnarray}
x(t)-x_i\!&=&\!\!-\frac{\alpha\,\Delta{Q}}{C_\mathrm{eq}}\,[1-e^{-(t-t_i-\Delta{t})/\tau}],
\label{Eq:x_later-times}
\\
T(t)-T_i\!&=&\!\!-\frac{\Delta{Q}}{C_\mathrm{eq}} \!\left(1+
\frac{k_i\alpha^2T_i e^{-(t-t_i-\Delta{t})/\tau}}{C_\mathrm{eq}-k_i\alpha^2T_i}\right)\!, \qquad \ \ 
\end{eqnarray}
for all $t> t_i +\Delta{t}$.
These quantities  only arrive at their expected equilibrium at $t\to\infty$, but they get exponentially close to equilibrium once  $t-t_i-\Delta{t} \gg \tau$.
Here the expected equilibrium temperature after extracting $\Delta{Q}$ of heat is 
$T=T_i-\Delta{Q}/C_\mathrm{eq}$, and then Eq.~(\ref{Eq:x_eq}) tells us that the equilibrium value of $x$ at $T$ 
is $x_i -\alpha\Delta{Q}/C_\mathrm{eq}$.

If instead of $x$ we have the field $\vec{A}(\vec{r})$, the description of the relaxation on this stage, though similar conceptually, would be more technically difficult. Expanding $\vec{A}(\vec{r})$ in the basis of eigenfunctions of the radial Laplacian, we would obtain an infinite set of ordinary differential equations for the amplitudes of these eigenfunctions with different relaxation times, instead of a single equation for~$x$ with a single relaxation time, $\tau$.
The dynamics of this infinite set of amplitudes under these equations are coupled through the equation for $U$. Finding the explicit trajectory in this case is beyond the scope of our short study.
However, as each component decays at a different rate,  it is clear that the spatial profile of the field will not have a fixed shape during the decay. 
This invalidates any ``fixed profile'' assumption, such as  
Eq.~(1) in Ref.~\cite{Hirsch}.

\section{Joule heating for full trajectory}
Now we have the full system dynamics, we can directly calculate the Joule heating.
The total Joule heating between time $t_i$ and $t_i+\Delta{t}$ (when the system is coupled to the reservoir)
is given by $x(t)$ in Eq.~(\ref{Eq:x_early-times}), so 
\begin{eqnarray}
\Delta{Q}_{J}(t_i+\Delta{t};t_i) 
\!\!&=&\!\! \eta_i \int_{t_i}^{t_i+\Delta{t}} \rmd t \left(\frac{\rmd x}{\rmd t}\right)^2   
\nonumber \\
&\approx&   \frac{\eta_i \alpha^2C_\mathrm{eq}}{ \dot Q \tau^2}\,\Delta T^3 + {\cal O}(\Delta T^4),\qquad
\label{Eq:Q_J-growth}
\end{eqnarray}
where we have used $\Qdot\Delta t = \Delta{Q}= C_\mathrm{eq} \Delta{T}$.

Next, the total Joule heating after time  $t_i+\Delta{t}$ (when the system relaxes back to equilibrium) is given by $x(t)$ in Eq.~(\ref{Eq:x_later-times}). 
Integrating this up to times large enough that it has decayed to close to equilibrium --- i.e. up to $t-(t_i+\Delta{t}) \gg \tau$ --- gives
\begin{eqnarray}
\Delta{Q}_J (\infty;t_i+\Delta{t})\!\!&\approx&\!\!\int \limits_{t_i+\Delta{t}}^\infty\frac{\eta_i}{\tau^2}\!
\left(\frac{\alpha\,\Delta{Q}}{C_\mathrm{eq}}\right)^{\!2} e^{-2(t-t_i-\Delta{t})/\tau}\,\rmd{t}
\nonumber\\
&=&\!\!\frac{\eta_i\alpha^2}{2\tau} \Delta{T}^2+ {\cal O}(\Delta T^3).
\label{Eq:Q_J-decay}
\end{eqnarray}
Thus protocol $(b)$'s total Joule heating $\Delta{Q}_J^{(b))}\sim {\cal O}(\Delta T^2)$
 for small $\Delta T$,  
because it is completely dominated by the decay to equilibrium,
so it is given by Eq.~(\ref{Eq:Q_J-decay}). 
Interestingly, this means that the Joule heating is independent of the magnitude of the damping $\eta_i$,
because the timescale for the decay to equilibrium $\tau \propto\eta_i$.  If the damping is small,
the Joule heating per unit time is small, but the decay to equilibrium takes a long time,  so the 
total Joule heat generated $\Delta Q_J$ is independent of $\eta_i$.

As we do not have the explicit trajectories for the situation where we replace $x$ by the field ${\bf A}({\bf r})$, we cannot give exact forms for the leading order contributions to the Joule heating, 
as in Eqs.~(\ref{Eq:Q_J-growth}), (\ref{Eq:Q_J-decay}).  However  it is not hard to see that  the evolution from $t_i$ to $t_i+\Delta{t}$ will have Joule heating of ${\cal O}(\Delta T^3)$.
The fact that the subsequent decay to equilibrium 
is controlled by a set of decay times that are independent of $\Delta{T}$, means that
it will have Joule heating of ${\cal O}(\Delta T^2)$.
Thus both contributions to Joule heating are of the same order as in the above calculation for $x$.

The main message of this section is that one has to go to ${\cal O}(\Delta T^2)$ to see the effect of Joule heating on the entropy,
it makes no contribution to the ${\cal O}(\Delta T)$ calculation in Ref.~\cite{Hirsch}.

\vskip 7mm
%===========================================
\centerline{{\bf PART 2: THERMODYNAMIC CONSISTENCY}}
\centerline{{\bf IN TIME-DEPENDENT GINZBURG-LANDAU}}
\vskip 3mm

In the section ``Cooling a superconductor'' we make a simplification of 
the Ginzburg-Landau free energy.  We eliminate the dynamics of the superconducting order parameter,  $\psi({\bf r})$, by assuming it always takes its zero-field equilibrium value. This means we only had to treat the dynamics of  ${\bf A}({\bf r})$, enabling us to make concrete calculations for fast cooling (in particular to find the magnitude of the Joule heating for small $\Delta{T}$).  This simplification is clearly wrong close to the superconducting transition,
where $\psi({\bf r})$ couples to ${\bf A}({\bf r})$. 
The simplest way to treat systems close to the transition (or going through the transition) would be to
use time-dependent Ginzburg-Landau theory for coupled dynamics of ${\bf A}({\bf r})$ and $\psi({\bf r})$,
see e.g. section 1.2 of Ref.~\cite{Kopnin}.

In this case, to disprove Refs.~\cite{Hirsch,Hirsch2}'s claims of thermodynamic inconsistency,
we can apply the logic in section ``Entropy change with fast cooling'' above to the 
 time-dependent Ginzburg-Landau theory.
The Ginzburg-Landau free energy density is
\begin{eqnarray}
{\cal F} ({\bf A},\psi)&=& {\cal F}_n \, +\, \alpha |\psi|^2 \,+\, {1\over 2} \beta |\psi|^4 
\nonumber \\
& & {}+ {1\over 2m} \big|(-i\hbar \nabla -2e{\bf A})\psi\big|^2
\nonumber \\
& & {}+ {1 \over 2\mu_0} \,\big|\nabla \times {\bf A}\big|^2,
\label{Eq:F}
\end{eqnarray}
where $\psi({\bf r})$ is the complex order parameter,  and ${\bf A}({\bf r})$ is the vector potential in the superconductor.  
The first term, $\mathcal{F}_n$, is the free-energy density of the normal phase in zero magnetic field.
Taking the functional derivatives of $F$ with respect to  ${\bf A}({\bf r})$  and $\psi({\bf r})$
will give their two conjugate forces;
let $f_1({\bf A},\psi)$ be the conjugate force of ${\bf A}({\bf r})$, and 
$f_2({\bf A},\psi)$ be the conjugate force of $\psi({\bf r})$.
These forces are zero at equilibrium, and push the displacements (${\bf A}({\bf r})$  and $\psi({\bf r})$) towards their equilibrium values.
The work done by these forces is
\begin{eqnarray}
\delta W\!&=&\!\int \rmd^3 {\bf r}\ \big[ f_1({\bf A},\psi)\, \delta{\bf A} \nonumber\\
&&{}+ f_2({\bf A},\psi)\, \delta\psi+ f_2^*({\bf A},\psi)\,\delta\psi^* \big]. \quad
\end{eqnarray}

Now we need equations which give the response of ${\bf A}({\bf r})$  and $\psi({\bf r})$ to the forces,
which will tell us how the system will relax to equilibrium.
One can imagine many mechanisms for the relaxation of both ${\bf A}({\bf r})$  and $\psi({\bf r})$.
The one proposed in our Eq.~(\ref{eq:Maxwell3}) ---based on the ideas in Ref.~\cite{Hirsch}---
induces simple  over-damped relaxation of ${\bf A}({\rm r})$ towards equilibrium.
Other relaxation mechanisms could easily exist, and at least one must exist to relax the order-parameter $\psi({\bf r})$.
These relaxation mechanisms could take many forms.
Rather than discuss all possibility,
let us take the simple example of time-dependent Ginzburg-Landau equations in Section 1.2 of \cite{Kopnin};
\begin{eqnarray} 
\label{Eq:SCeta1}
\eta_1({\bf A},\psi) \,\frac{\rmd {\bf A}}{\rmd t} &=& f_1({\bf A},\psi),
\\
\label{Eq:SCeta2}
\eta_2({\bf A},\psi) \,\frac{\rmd \psi}{\rmd t} &=&  f_2^*({\bf A},\psi),
\end{eqnarray}
so both ${\bf A}({\rm r})$ and $\psi({\rm r})$ exhibit simple overdamped dynamics. 
Other cases are discussed in the next section.

The internal heating generated by the damping process will then take the form
\begin{eqnarray}
\frac{\rmd Q_J}{\rmd t}
\!&=&\! \int \rmd^3 {\bf r}\, \left(
\eta_1\,  \left(\frac{\rmd {\bf A}}{\rmd t}\right)^{\!2} + 
 2\eta_2 \, \left|\frac{\rmd \psi}{\rmd t}\right|^{\!2} \right), \ \ \qquad
\label{Eq:Q_J-GL1}
\end{eqnarray}
where we use the subscript ``$J$'' to make the analogy with Joule heating, even if the internal heating 
could be of a very different nature.
Thus the equivalent of our  Eq.~(\ref{Eq:DS^b}) for the fast cooling protocol $(b)$, 
gives the entropy change of the universe after time $t_f \geq t_i+\Delta{t}$ as
\begin{eqnarray}
\Delta S_{univ}^{(b)} = {\Delta Q \over T_2}-\int_{t_i}^{t_i+\Delta{t}} {\rmd t \over T}\, \Qdot + 
\int_{t_i}^{t_f} {\rmd t \over T}\, 
\left(\frac{\rmd Q_J}{\rmd t}\right). \ \ \
\label{Eq:DS^b-GL}
\end{eqnarray}
As with our  Eq.~(\ref{Eq:DS^b}), we use Eq.~(\ref{Eq:dU}) to write $\rmd{U}/\rmd{t}=-\Qdot$, and use Eq.~(\ref{Eq:SCeta1}), (\ref{Eq:SCeta2}) to eliminate the $\eta$s.
Then
\begin{eqnarray}
\Delta S_{univ}^{(b)} \!\!&=&\!\! {\Delta Q \over T_2}+
\int_{t_i}^{t_f} \!{\rmd t \over T}\, \left[\frac{\rmd U}{\rmd t}\right.
\nonumber \\
& &{}+ \int \!\rmd ^3{\bf r}\left.\left( f_1 \frac{\rmd {\bf A}}{\rmd t}
+ f_2  \frac{\rmd \psi}{\rmd t}
 + f_2^*  \frac{\rmd \psi^*}{\rmd t}
\right)\right]\!. \qquad \ \ \ 
\label{Eq:DS^b-GL-2ndresult}
\end{eqnarray}
As with our Eq.~(\ref{Eq:DS^b-2ndresult}), the fundamental thermodynamic relation tells the integrand of the $t$-integral is $\rmd{S}/\rmd{t}$, so the integral is $S(x_f,T_f)-S(x_i,T_i)$.
Thus $\Delta S_{univ}^{(b)}$ depends on the initial and final states, but not on the trajectory between these states.     This confirms that the time-dependent  Ginzburg-Landau theory does not exhibit the thermodynamic inconsistent claimed in Refs.~\cite{Hirsch,Hirsch2}.

\section{More complicated dynamics}
The introduction of Ref.~\cite{Kopnin} explains the limitations of the above time-dependent  Ginzburg-Landau model;
indeed that book's objective was to provide better microscopic models of superconductors.  In general,
improving the model means adding the microscopic dynamics of more degrees-of-freedom (quasi-particles, phonons, etc), instead of just assuming they always take their equilibrium value.
Above we have disproved Refs.~\cite{Hirsch,Hirsch2}'s claims of thermodynamic inconsistency 
for (i) a model with dynamics of a single parameter $x$,  (ii) a model with dynamics of a field ${\bf A}({\bf r})$, and (iii) a model with dynamics of a pair of fields ${\bf A}({\bf r})$ and $\psi({\bf r})$.
Thus it seems implausible that simply adding the dynamics of more degrees-of-freedom will lead to a violation of thermodynamic consistency.

%=================================================
%=================================================
%=================================================
%=================================================

\end{document}